\documentclass{biometrika}

\usepackage{amsmath}

\usepackage{newtxtext}
\usepackage[subscriptcorrection]{newtxmath}
\usepackage{dsfont}

\def\PP{{\rm pr}}
\def\E{E}
\def\Var{{\rm var}}
\def\ind{\mathds{1}}
\def\eqd{\stackrel{\rm d}{=}}
\newcommand{\A}{\ensuremath{\mathcal{A}}}
\newcommand{\X}{\ensuremath{\mathcal{X}}}
\newcommand{\G}{\ensuremath{\mathbb{G}}}
\newcommand{\R}{\ensuremath{\mathbb{R}}}
\newcommand{\Omd}{\ensuremath{\Omega_{\rm d}}}
\newcommand{\Omc}{\ensuremath{\Omega_{\rm c}}}

\makeatletter
\let\newthanks\thanks
\makeatother

\makeatletter
\def\ps@bottomfolio{%
  \def\@oddfoot{\hfil\thepage\hfil}\let\@evenfoot\@oddfoot
  \def\@oddhead{}\def\@evenhead{}\let\@mkboth\markboth}
\makeatother

\begin{document}



\markboth{D. Dutz and X. Zhang}{Limitations of randomization tests}

\title{Limitations of randomization tests in finite samples\thanks{We thank
Xinran Li, Guillaume Pouliot, Azeem Shaikh, and Max Tabord-Meehan for helpful
discussions.}}

\author{DENIZ DUTZ}
\affil{Department of Economics, University of Chicago
\email{ddutz@uchicago.edu}}

\author{\and XINYI ZHANG}
\affil{Department of Mathematics, Columbia University
\email{xz3272@columbia.edu}}

\maketitle
\pagestyle{bottomfolio}
\thispagestyle{bottomfolio}

\begin{center}
August 14, 2026
\end{center}

\begin{abstract}
Randomization tests deliver exact finite-sample type I error control when the null satisfies the randomization hypothesis. In practice, these guarantees often require invariance conditions stronger than the null of primary interest. For example, sign-change tests of mean zero are exact only under symmetry and may over-reject for non-symmetric mean-zero distributions. We investigate whether this mismatch reflects the use of particular transformations or a more fundamental limitation. To do so, we provide a simple necessary and sufficient condition for a null hypothesis to admit a randomization test. Applying it to one-sample problems, we characterize the nulls that do and do not admit randomization tests, and show that several common nulls, including mean zero, admit none. Among tests using linear group actions, for instance, the admissible nulls are limited to subsets of symmetric or Gaussian distributions. When a null does admit a randomization test, our condition identifies transformations that implement it. These results confirm that the absence of exact finite-sample validity is inherent for many commonly studied nulls and that practitioners using existing tests are not forgoing feasible exact alternatives.
\end{abstract}


\section{Introduction}

Consider the problem of testing a null hypothesis given a finite sample of data.
If this null satisfies a group invariance property referred to as the `randomization hypothesis', then one can construct randomization tests that obtain exact finite-sample type I error control.
Intuitively, this property requires that, under the null, certain transformations of the data leave its distribution unchanged.
In part due to the appeal of finite-sample validity, there is a large methodological literature on randomization tests and a similarly large applied literature using these methods \citep[see the discussions in, e.g.,][]{zhang2023randomization,ritzwoller2024randomization}.

In practice, randomization tests often rely on invariance conditions that are stronger than the null hypothesis of primary interest.
This gap undermines the guarantee: the test can over-reject when the data are drawn from a distribution that satisfies the hypothesis of primary interest but not the stronger conditions.
For example, a commonly used randomization test for the null of mean zero with i.i.d. observations is based on sign changes.
This test controls finite-sample size only under symmetry and may over-reject for non-symmetric mean-zero distributions.

This paper investigates whether the failure to control type I error for certain null hypotheses, for example the null of mean zero, stems from the specific tests used in practice, such as sign changes, or from a more fundamental limitation.
To formalize this distinction, we provide a simple necessary and sufficient condition, involving no group-theoretic structure, for a null to satisfy the randomization hypothesis.

We apply this condition to one-sample problems.
We show that certain null hypotheses, such as the null of mean zero, do not admit randomization tests that achieve exact finite-sample validity.
These results relate to classical non-existence results for nonparametric inference \citep{bahadur1956nonexistence,lehmann1990pointwise,romano2004non}.
By focusing on randomization tests, we derive conditions for the existence and non-existence of exact finite-sample procedures across a range of null hypotheses and distributional classes.

Our results are also constructive.
On finite supports, we give an explicit characterization of the nulls that admit randomization tests and show how our condition identifies transformations that implement them. On continuous supports, we characterize the tests available from linear group actions and show that they are limited: the admissible nulls are the symmetric distributions or subsets of Gaussian distributions. 
Along the way, we provide explicit constructions of tests that may be of theoretical or applied interest.

Taken together, these results yield a positive implication for applied work: the use of standard randomization tests does not overlook alternative procedures that achieve exact finite-sample validity for many commonly studied nulls.
They also affirm and motivate a focus on asymptotic properties of randomization tests \citep[see, e.g.,][]{romano1989bootstrap,romano1990behavior,diciccio2017robust,canay2017randomization,lei2021assumption,pouliot2024ttest,bai2024inference}.
Finally, our results suggest that one-sample randomization tests that go beyond symmetry or normality must rely on non-linear transformations.


\section{The randomization test}\label{sec2}

\subsection{Construction of the randomization test}

The construction follows \citet{hoeffding1952large,lehmannromano}.
Let $X = (X_1,\ldots,X_n)$ be data taking values in $\X$ with distribution $P \in \Omega$. Consider the problem of testing $H_0 : P \in \Omega_0$, and suppose that there is a group $\G$ of measurable transformations $g:\X\to\X$ such that $gX \eqd X$ whenever $X \sim P \in \Omega_0$. For ease of exposition, we assume $\G$ is finite with size $M$; see Remark 2.1 of \citet{ritzwoller2024randomization} for one extension to infinite groups.

Let $T:\X\to\R$ be any test statistic. For every $x \in \X$, let
$T^{(1)}(x) \leq \cdots \leq T^{(M)}(x)$ be the ordered values of $T(gx)$ as
$g$ ranges over $\G$.
Given nominal level $\alpha \in (0,1)$, let $k \equiv M - \lfloor M\alpha \rfloor$, where $\lfloor M\alpha \rfloor$ denotes the largest integer less than or equal to $M\alpha$. Let $M^{+}(x)$ and $M^{0}(x)$ be the number of values $T^{(j)}(x)$ that are greater than $T^{(k)}(x)$ and equal to $T^{(k)}(x)$, respectively. Set
\[
    a(x) = \frac{M \alpha - M^{+}(x)}{M^{0}(x)}.
\]
For $x,y \in \X$, define
\[
    \phi(y;x) \equiv
    \begin{cases}
        1, & T(y) > T^{(k)}(x), \\
        a(x), & T(y) = T^{(k)}(x), \\
        0, & T(y) < T^{(k)}(x).
    \end{cases}
\]
The randomization test is $\phi^\star(x) \equiv \phi(x;x)$.
The following theorem shows that the randomization test has nominal level $\alpha$.
Our proof highlights the role of the group structure; see Remark \ref{rm:group-structure} below.

\begin{theorem}\label{thm:test}
    Suppose that for any $g \in \G$, $gX \eqd X$ whenever $X \sim P \in \Omega_0$. Then $\E_P(\phi^\star(X)) = \alpha$ for every $P \in \Omega_0$.
\end{theorem}

\begin{proof}[of Theorem~\ref{thm:test}]
    Consider any $X \sim P \in \Omega_0$. Since $gX \eqd X$ for each $g \in \G$, $\E_P(\phi^\star(X)) = \E_P(\phi^\star(gX))$, and so
    \[
        \E_P(\phi^\star(X))
        = M^{-1} \sum_{g \in \G} \E_P(\phi^\star(gX))
        = \E_P \left( M^{-1} \sum_{g \in \G} \phi(gX;gX) \right).
    \]
    Fix $x \in \X$.
    Since the values $T(gx)$, $g \in \G$, are exactly $T^{(1)}(x),\ldots,T^{(M)}(x)$, the construction of $a(x)$ gives $M^{-1} \sum_{g \in \G} \phi(gx;x) = \alpha$.
    The result then follows from showing
    \begin{equation}\label{eq:thmtest1}
        M^{-1} \sum_{g \in \G} \phi(gx;gx) = M^{-1} \sum_{g \in \G} \phi(gx;x).
    \end{equation}
    Since $\G$ is a group, $\G g = \G$, so the values $T(h(gx)) =
    T((hg)x)$, $h \in \G$, are a rearrangement of the values $T(hx)$,
    $h \in \G$. Equivalently, $T^{(j)}(gx) = T^{(j)}(x)$ for every $j$.
    It follows that $\phi(gx;gx) = \phi(gx;x)$, which yields \eqref{eq:thmtest1} and thus the result.
\end{proof}

\begin{remark}\label{rm:group-structure}
    Suppose that we did not require that the collection of invertible transformations $\G$ is a group.
    For \eqref{eq:thmtest1} to hold for every test statistic, we require that $\phi(gx;gx) = \phi(gx;x)$ for every $g \in \G$, which is the requirement that $\G g = \G$ for all $g \in \G$.
    A basic property of groups is that if the set $\G$ is a collection of invertible transformations, then $\G g = \G$ for all $g \in \G$ if and only if $\G$ is a group.
    In this sense, the requirement that $\G$ is a group is both necessary and sufficient.
\end{remark}

\subsection{Characterizing the randomization hypothesis}

We begin by defining the randomization hypothesis.

\begin{definition}[Randomization hypothesis]\label{def:randhyp}
    We say $\Omega_0$ satisfies the randomization hypothesis if there exists a group $\G$ of measurable transformations $g:\X\to\X$ such that
    \begin{equation}\label{eq:Omega_subset}
        \Omega_0 \subseteq \Omega_{\G} \subset \Omega,
    \end{equation}
    where $\Omega_{\G} \equiv \{P \in \Omega : g X \eqd X \text{ for all } g \in \G,\ X \sim P\}$.
\end{definition}

The set inclusions in \eqref{eq:Omega_subset} enforce that $\Omega_0$ is invariant to $\G$ while also ensuring that $\G$ is not invariant to all distributions in $\Omega$. The latter condition rules out trivial groups, such as the identity, that are invariant to all distributions.

Our analysis will focus on characterizing which null hypotheses do and do not satisfy the randomization hypothesis. For those that do, we will be interested in constructing a group to implement the test.
The following result provides a tool to answer both of these questions in a way that avoids group-theoretic concepts.

\begin{proposition}\label{prop:tool}
    The null $\Omega_0$ satisfies the randomization hypothesis if and only if there exists a bijective measurable function $f: \X \to \X$ such that
    \begin{equation}\label{eq:Omega_subsetf}
        \Omega_0 \subseteq \Omega_f \subset \Omega,
    \end{equation}
    where $\Omega_f \equiv \{P \in \Omega : f X \eqd X,\ X \sim P\}$. When this holds, a randomization test can be constructed using the group generated by $f$.
\end{proposition}

\noindent All omitted proofs can be found in Appendix \ref{app:proofs}. The proof of the forward direction is immediate and the proof of the reverse direction follows from showing that if $f$ is invariant to all distributions in $\Omega_0$, then so is $f^{-1}$. We can then show that $\Omega_0$ satisfies the randomization hypothesis using the group generated by $f$.

Proposition~\ref{prop:tool} shows that to study which null hypotheses satisfy the randomization hypothesis, it suffices to consider individual bijective functions.
Given such a function satisfying \eqref{eq:Omega_subsetf}, a randomization test can be implemented using the group it generates.
In the remainder of the paper, we use this result to study one-sample problems.

\section{One-sample tests on a finite support}\label{sec3}

In this section, we suppose the $X_i$ are drawn i.i.d. from $P$ supported on a finite set $\A \equiv \{\alpha_1,\ldots,\alpha_K\} \subset \R$, so that $X = (X_1,\ldots,X_n)$ takes values in $\X = \A^n$ with distribution the $n$-fold product of $P$.
We represent $P$ by its probability mass function $p: \A \to [0,1]$ and define
\[
    \Omd \equiv \left\{p : \A \to [0,1] \text{ such that } \sum_{k=1}^K p(\alpha_k) = 1\right\}.
\]
We identify each $P$ with its probability mass function $p$ and use the two interchangeably.

By Proposition~\ref{prop:tool}, a null hypothesis $\Omega_0 \subset \Omd$ satisfies the randomization hypothesis if and only if there exists a bijection $f: \A^n \to \A^n$ such that $\Omega_0 \subseteq \Omega_f \subset \Omd$. On a finite support, we can explicitly write $\Omega_f$ as
\begin{equation}\label{eq:discrete_form}
    \Omega_f = \bigcap_{x \in \A^n} \left\{ p \in \Omd : \prod_{j=1}^n p(x_j) = \prod_{j=1}^n p(f_j(x)) \right\}.
\end{equation}
Each set in this intersection is itself the invariance set of the bijection that swaps $x$ and $f(x)$ and fixes all other points.
This provides an explicit characterization of the condition in Proposition~\ref{prop:tool} for one-sample tests over finite supports.

\begin{proposition}\label{prop:tool_finite}
    The null $\Omega_0 \subset \Omd$ satisfies the randomization hypothesis if and only if there exist $x,y \in \A^n$ such that $y$ is not a permutation of $x$ and such that $\prod_{j=1}^n p(x_j) = \prod_{j=1}^n p(y_j)$ for all $p \in \Omega_0$.
\end{proposition}

\noindent Proposition~\ref{prop:tool_finite} is useful for two reasons.
First, it can be used to identify bijective functions that satisfy the invariance. We can then construct randomization tests by taking the groups generated by the functions. To illustrate this, Examples \ref{ex:construct_discrete_1} and \ref{ex:construct_discrete_2} of Appendix \ref{app:examples} use the result to construct the usual test of symmetry around zero and to construct a test of two points having equal mass.

Second, Proposition~\ref{prop:tool_finite} shows that null hypotheses satisfying the randomization hypothesis are characterized by equal probability mass assigned to distinct configurations, accounting for multiplicities.
Null hypotheses not defined by such restrictions therefore do not satisfy the randomization hypothesis.
In what follows, we apply Proposition~\ref{prop:tool_finite} to show that null hypotheses defined by $t$-th moments, including the mean, and by quantiles do not satisfy the randomization hypothesis.

\begin{proposition}\label{prop:discrete_moments}
    If $|\A| > 4$, then for any positive integer $t$ and any
    $\beta \in (\min_k \alpha^{t}_k, \max_k \alpha^{t}_k)$,
    $\Omega_0 = \{P \in \Omd : \E_P(X_i^{t}) = \beta\}$ does not satisfy
    the randomization hypothesis.
\end{proposition}

\begin{proposition}\label{prop:discrete_quantile}
    If $|\A| > 2$, then for any $p \in (0,1)$ and $q \in (\alpha_1,\alpha_K)$, $\Omega_0 = \{P \in \Omd : \PP(X_i \leq q) = p\}$ does not satisfy the randomization hypothesis.
\end{proposition}

\noindent The proofs of Propositions~\ref{prop:discrete_moments} and~\ref{prop:discrete_quantile} proceed by showing that for any $x,y \in \A^n$ that are not equal up to permutation, one can construct a distribution in the null that assigns different probabilities to $x$ and $y$.
The result then follows from Proposition~\ref{prop:tool_finite}.

\begin{remark}
    The cardinality requirements in Propositions~\ref{prop:discrete_moments} and~\ref{prop:discrete_quantile} exclude edge cases in which the null does satisfy the randomization hypothesis. For example, when $\A = \{-L,0,L\}$ for some $L > 0$, the null of mean zero coincides with symmetry, which satisfies the randomization hypothesis via sign changes.
    The proof of Proposition~\ref{prop:discrete_moments} provides an analogous edge case for $|\A| = 4$.
    For Proposition~\ref{prop:discrete_quantile}, the result continues to hold when $|\A| = 2$ except at $p = 1/2$, where the transformation swapping the two points yields a valid randomization test.
\end{remark}

\section{One-sample tests on a continuous support}\label{sec4}

In this section, we suppose $X_i$ is drawn i.i.d. from $P$ with a density on the real line.
In what follows, we first consider groups of linear transformations, and explicitly characterize the set of all null hypotheses that are invariant to any such group. 
We then consider arbitrary groups, and develop and apply a necessary condition for a null hypothesis to satisfy the randomization hypothesis.

\subsection{Groups of linear transformations}

A group of linear transformations $\G$ is one in which each element $g$ corresponds to a linear map $A_g:\R^n \to \R^n$.
The following result characterizes the distributions invariant to such groups.

\begin{proposition}\label{prop:linear}
    Let $\Omc$ be a set of distributions with densities on the real line, and let $\G$ be a group of linear transformations acting on $\R^n$. Then one of the following holds: (i) $\Omega_{\G} = \Omc$; (ii) $\Omega_{\G}$ is the set of distributions in $\Omc$ that are symmetric about zero; (iii) $\Omega_{\G} \subseteq \{N(\mu,\sigma^2) : \mu \in \R,\ \sigma^2 > 0\}$; or (iv) $\Omega_{\G} = \emptyset$.
\end{proposition}

The proof proceeds by writing $gX = AX$ for some matrix $A$.
If $A$ is a signed permutation matrix, invariance either imposes no restriction or requires symmetry.
Otherwise, some component depends linearly on multiple coordinates, and the Darmois--Skitovich theorem implies the distribution belongs to the Gaussian family.

Proposition~\ref{prop:linear} shows that, when restricted to linear group actions, one-sample randomization tests admit very few invariances beyond permutations: they can only test membership in the class of distributions symmetric about zero or in subclasses of the Gaussian family.
The former is implemented by sign changes, while Examples \ref{ex:construct_cts_1} and \ref{ex:construct_cts_2} in Appendix \ref{app:examples} illustrate the latter.
The result rules out using rotations or other nontrivial linear transformations to construct randomization tests for distributional classes beyond these two.

\subsection{Arbitrary groups}\label{sec:arb_cts}

We now allow $\G$ to be arbitrary.
Classical non-existence results of \citet{bahadur1956nonexistence} and \citet{romano2004non} show that sufficiently rich classes of distributions cannot be tested in a nonparametric setting.
We show that a related conclusion holds in our context under a local version of their richness condition.

Let $\Omc(\mathcal{B})$ be the set of distributions with a density supported on $\mathcal{B}\subseteq \R$.
Informally, a null $\Omega_0 \subset \Omc(\mathcal{B})$ is locally dense if it can locally replicate any density on finitely many small intervals, up to scale.

\begin{definition}[Locally dense]\label{def:locally-dense}
A set $\Omega_0 \subset \Omc(\mathcal{B})$ is locally dense if there exists $L:\mathbb{N} \to \R_+$ such that for any collection of intervals $I_1,\ldots,I_m$ with $|I_i| < L(m)$ and any density $p$ supported on $\bigcup_{i=1}^m I_i$, there exist $H \in \Omega_0$ with density $h$ and $\alpha \in (0,1]$ such that $h(x) = \alpha p(x)$ for all $x \in \bigcup_{i=1}^m I_i$.
\end{definition}

\noindent The following result shows that locally dense null hypotheses do not satisfy the randomization hypothesis.

\begin{proposition}\label{prop:continuous}
    If $\Omega_0$ is locally dense, then $\Omega_0$ does not satisfy the randomization hypothesis.
\end{proposition}

The proof uses Proposition~\ref{prop:tool} and shows that for any function $f$ with $\Omega_0 \subseteq \Omega_f$, local denseness implies $\Omega_f = \Omc(\mathcal{B})$.
Intuitively, local denseness ensures that the null can approximate any density on finitely many regions, allowing the invariance to extend to all distributions.

We now use Proposition~\ref{prop:continuous} to show that the same negative results obtained in Section \ref{sec3} also hold when considering continuous distributions or when considering continuous distributions over a fixed interval.

\begin{proposition}\label{prop:cts_case1}
    Let $t$ be a positive integer, and let $\beta \in \R$ if $t$ is odd and $\beta > 0$ if $t$ is even. Then $\Omega_0 = \{P \in \Omc(\R) : \E_P(X_i^{t}) = \beta\}$ is locally dense and thus does not satisfy the randomization hypothesis.
\end{proposition}

To illustrate the content of this result, consider the null $\E_P(X_i) = \beta$.
Let $I_1,\ldots,I_m$ be intervals, each of width less than $L(m)$.
For any density $p$ supported on these intervals with mean $\gamma_p$ and any $\alpha \in (0,1]$, construct a distribution $H$ with density $\alpha p(x)$ on these intervals.
Then $H$ has mean
\[
\alpha \gamma_p + (1-\alpha)\E_H(X_i \mid X_i \notin \cup_i I_i).
\]
By placing the remaining mass on two intervals above and below $\beta$, one can choose $\alpha$ and the remaining mass so that the mean equals $\beta$.
Thus the null is locally dense.
The following result is shown using a similar construction.

\begin{proposition}\label{prop:cts_case2}
    For any $p \in (0,1)$ and $q \in \R$, $\Omega_0 = \{P \in \Omc(\R): \PP(X_i \leq q) = p\}$ is locally dense and thus does not satisfy the randomization hypothesis.
\end{proposition}

\begin{remark}\label{rm:extend-cts}
We show in Appendix \ref{app:proofs3} that Proposition~\ref{prop:cts_case1} extends to $\Omc([b_0,b_1])$ for any interval $[b_0,b_1]$ and any $\beta$ in the interior of the set of attainable $t$-th moments.
The same extension holds for Proposition~\ref{prop:cts_case2} for any $q \in (b_0,b_1)$.
We also show that the null hypothesis of fixed variance is locally dense and therefore does not satisfy the randomization hypothesis.
\end{remark}

\section{Conclusion}\label{sec:conclusion}

Randomization tests deliver exact finite-sample type I error control when the null satisfies the randomization hypothesis, but in practice they often rely on invariance conditions stronger than the null of primary interest.
This paper develops a simple characterization of the nulls that admit randomization tests and applies it to one-sample problems.

The results are largely negative.
Many standard nulls, including mean zero, admit no randomization test with exact finite-sample validity, and tests based on linear group actions can only detect membership in the symmetric or Gaussian families.
Testing beyond symmetry or normality therefore requires non-linear transformations, and even these cannot deliver exact tests for many commonly studied nulls. At the same time, the characterization is constructive: when a null does admit a randomization test, it identifies transformations that implement one.

Overall, our findings indicate that current practice is not forgoing feasible exact procedures, and they further motivate the study of asymptotic validity of randomization tests.

\bibliographystyle{biometrika}
\bibliography{ref}

\clearpage

\clearpage
\setcounter{page}{1}
\begin{center}
{\titlefont Supplemental material\par}
\end{center}
\vskip20pt

\appendix

\appendixone
\section{Examples of randomization tests}\label{app:examples}

\begin{example}[Randomization test of symmetry over discrete support]\label{ex:construct_discrete_1}
For $L>0$, let $\A \equiv \{-L,\ldots,0,\ldots,L\}$ and let $\Omega_0$ be the set of distributions that are symmetric around zero. Consider any $x,y$ with $x_j \neq 0$ for some $j$ and $y=-x$. We then have that $c(x_j;x) > c(x_j;y)$ and that for all $p \in \Omega_0$, $\prod_{j=1}^n p(x_j) = \prod_{j=1}^n p(-x_j) = \prod_{j=1}^n p(y_j)$. By Proposition~\ref{prop:tool_finite}, $\Omega_0$ satisfies the randomization hypothesis. We can then construct a randomization test for $\Omega_0$ by taking the group arising from any collection of functions that send some values of $x \in \A^n$ to $-x$ and otherwise act as the identity function.
\end{example}

\begin{example}[Randomization test of equal mass over discrete support]\label{ex:construct_discrete_2}
For any $\mathcal{A}$, let $\Omega_0$ be the set of distributions where $p(\alpha_\ell) = p(\alpha_{\ell'})$ for some $\alpha_\ell \neq \alpha_{\ell'}$. Consider any $x,y$ with $x$ containing $\alpha_{\ell}$ but not $\alpha_{\ell'}$ and $y_j = \alpha_{\ell'}\ind[x_j = \alpha_\ell] + \alpha_{\ell}\ind[x_j = \alpha_{\ell'}] + x_j \ind[x_j \neq \alpha_\ell, x_j \neq \alpha_{\ell'}]$. We then have that $c(\alpha_{\ell};x) > c(\alpha_{\ell};y)$ and for all $p \in \Omega_0$, $\prod_{j=1}^n p(x_j) = \prod_{j=1}^n p(y_j)$. By Proposition \ref{prop:tool_finite} $\Omd$ satisfies the randomization test. We can then construct a randomization test for $\Omega_0$ by taking the group arising from any collection of functions that acts as the identity except for swapping the values of $\alpha_{\ell}$ and $\alpha_{\ell'}$.
\end{example}

\begin{example}[Randomization test for the mean-zero Gaussian family over continuous support] \label{ex:construct_cts_1}
    Let $\G$ be the orthogonal group of degree $n$. By Maxwell's Theorem, the set of invariances given $\G$ is $\Omega_{\G} = \{N(0,\sigma^2) : \sigma^2 >0\}$. Thus, this group can be used to construct a randomization test of a distribution being Gaussian with mean zero.
\end{example}

\begin{example}[Randomization test for the Gaussian family over continuous support] \label{ex:construct_cts_2}
    Let $\G$ be the set of transformations generated by $3\times 3$ block permutations of the matrix
    \[
    A \equiv
    \begin{pmatrix}
    \frac{2}{3} & -\frac{1}{3} & \frac{2}{3}\\
    \frac{2}{3} & \frac{2}{3} & -\frac{1}{3} \\
    -\frac{1}{3} & \frac{2}{3}& \frac{2}{3}\\
    \end{pmatrix}.
    \]
    Let $X_1, X_2, X_3$ be i.i.d. draws from $N(\mu, \sigma^2)$ for some $\mu, \sigma^2 >0$, and set $X = (X_1, X_2, X_3)$ and $V = (\mu, \mu, \mu)$. Since $A$ is orthogonal, $A(X-V) \eqd X-V$, and since $AV = V$ this gives $AX \eqd X$. Inductively $A^kX \eqd X$ for all $k \in \mathbb{Z}$, so $\{N(\mu, \sigma^2): \mu, \sigma^2 >0\} \subseteq \Omega_{\G}$.
    The inclusion $\Omega_{\G} \subset \Omc$ is strict: if the $X_i$ are i.i.d. uniform on $(0,1)$, the first coordinate of $AX$ is supported on $(-1/3,4/3)$, so $AX \not\eqd X$.
    Provided $n \geq 3$, the group $\G$ can thus be used to construct a randomization test of a distribution being Gaussian.
\end{example}

\appendixtwo
\section{Proofs}\label{app:proofs}

Appendix \ref{app:proofs1} provides proofs of all propositions except Proposition \ref{prop:continuous}.
The proof of Proposition \ref{prop:continuous} requires introducing additional notation and is developed in Appendix \ref{app:proofs2}. 
Appendix \ref{app:proofs3} states and proves the claims made in Remark \ref{rm:extend-cts}.

\subsection{Proofs of Propositions \ref{prop:tool}--\ref{prop:cts_case2}, excluding Proposition \ref{prop:continuous}}\label{app:proofs1}

\begin{proof}[of Proposition~\ref{prop:tool}]
    The forward direction follows from noting that $\Omega_{\G} = \cap_{g \in \G} \Omega_g$, and since \eqref{eq:Omega_subset} holds, there must be at least some $f \in \G$ such that \eqref{eq:Omega_subsetf} holds. Group elements have inverses and thus $f$ is bijective.

    For the reverse direction, consider a bijective measurable $f$ satisfying \eqref{eq:Omega_subsetf}. The set generated by $f$, denoted $\langle f \rangle \equiv \{f^K : K \in \mathbb{Z}\}$, is a group with composition as the operation, with $f^{-K} = (f^{-1})^{K}$ where $f^{-1}$ denotes the inverse of $f$. Then $\Omega_{\langle f \rangle} \subseteq \Omega_f \subset \Omega$.

    It remains to show that $\Omega_0 \subseteq \Omega_{\langle f \rangle}$, which is equivalent to showing that $\PP(f^K X \in A) = \PP(X \in A)$ for all $X \sim P \in \Omega_0$, $K \in \mathbb{Z}$, and measurable $A$. When $K=0$, the proof follows trivially since the identity is invariant to all distributions. For any integer $K > 0$,
    \[
        \PP(f^K X \in A) = \PP(fX \in f^{K-1}A) = \PP(X \in f^{K-1}A) = \PP(f^{K-1}X \in A),
    \]
    where we used that $f$ is invertible and also used that $f$ is measurable so that $f^{K-1}A$ is measurable and thus, since $\Omega_0 \subseteq \Omega_f$, we have $\PP(fX \in f^{K-1}A) = \PP(X \in f^{K-1}A)$. Repeating this step, we obtain $\PP(f^K X \in A) = \PP(X \in A)$. Lastly,
    \[
        \PP(f^{-1}X \in A) = \PP(X \in fA) = \PP(fX \in fA) = \PP(X \in A),
    \]
    where the second equality used that $\Omega_0 \subseteq \Omega_f$ and that $fA$ is measurable so that $\PP(X \in fA) = \PP(fX \in fA)$. Then for $K>0$, arguments as above imply
    \[
        \PP(f^{-K} X \in A) = \PP(f^{-1} X \in f^{-(K-1)}A) = \PP(X \in f^{-(K-1)}A) = \PP(f^{-(K-1)}X \in A).
    \]
    Repeating this step, we obtain $\PP(f^{-K} X \in A) = \PP(X \in A)$, as required.
\end{proof}

\begin{proof}[of Proposition~\ref{prop:tool_finite}]
    By Proposition~\ref{prop:tool}, $\Omega_0$ satisfies the randomization hypothesis if and only if there exists a bijective $f$ such that $\Omega_0 \subseteq \Omega_f \subset \Omd$, where we drop measurability since all functions are measurable over discrete sets. Expanding the form of $\Omega_f$ using \eqref{eq:discrete_form}, we have that
    \[
        \Omega_f = \bigcap_{x \in \A^n} \left\{ p \in \Omd : \prod_{j=1}^n p(x_j) = \prod_{j=1}^n p(y_j) \ \text{such that} \ y = f(x) \right\}.
    \]
    Then any $f$ such that $\Omega_0 \subseteq \Omega_f \subset \Omd$ must have some $x^\star \in \A^n$ and $y^\star = f(x^\star)$ such that
    \[
        \Omega_0 \subseteq \underbrace{\left\{ p \in \Omd : \prod_{j=1}^n p(x^\star_j) = \prod_{j=1}^n p(y^\star_j)\right\}}_{\equiv \Omega^\star} \subset \Omd.
    \]
    If $x^\star$ and $y^\star$ are permutations of each other so that $y^\star = (x^\star_{\sigma(1)}, \ldots, x^\star_{\sigma(n)})$ for some permutation $\sigma \in S_n$, then $\prod_{j=1}^n p(x^\star_j) = \prod_{j=1}^n p(y^\star_j)$ for all $p \in \Omd$, so $\Omega^\star = \Omd$. It thus must be that these two vectors are not permutations of each other. The proof is completed by showing that when this is not the case, there is some $p \in \Omd$ that does not belong to $\Omega^\star$.
    We can construct such a $p$ by noting that there is at least some element $\alpha \in \A$ that appears more times in $x^\star$ than in $y^\star$, and we can then define $p$ as placing mass $\epsilon_0$ on all values except $\alpha$ and mass $\epsilon_1 \neq \epsilon_0$ on $\alpha$, so that $\prod_{j=1}^n p(x^\star_j) \neq \prod_{j=1}^n p(y^\star_j)$, as required.
\end{proof}

\begin{proof}[of Proposition~\ref{prop:discrete_moments}]
    Define $c(\alpha; x)$ as the number of times the value $\alpha$ shows up in the vector $x$. To show that $\Omega_0 = \{p \in \Omd: \E_p(X_i^t) = \beta\}$ does not satisfy the randomization hypothesis, it suffices to show that for all $x,y \in \A^n$ with $c(\alpha;x) > c(\alpha;y)$ for some $\alpha \in \A$, there exists some $p \in \Omega_0$ such that $\prod_{j=1}^n p(x_j) \neq \prod_{j=1}^n p(y_j)$.

    Let $x,y \in \A^n$ be arbitrary such that $c(\alpha;x) > c(\alpha;y)$ for some $\alpha \in \A$.

    First, consider the case where $t$ is odd. For simplicity, let us assume that $\alpha_1 < \alpha_2 < \cdots < \alpha_K$. For $x,y \in \A^n$, we define $\beta_i(x,y) = c(\alpha_i;x) - c(\alpha_i;y)$ and notice that $\sum_{i=1}^K \beta_i(x,y) = 0$. Let us consider the following two cases.
    \begin{enumerate}
        \item There exists some $1 \leq i \leq K-1$ such that $\alpha_i^t < \beta < \alpha^t_{i+1}$. Consider the following subcases.
        \begin{enumerate}
            \item There exists some $i_1 \leq i$ and $i_2 \geq i+1$ such that $|\beta_{i_1}(x,y)| \neq |\beta_{i_2}(x,y)|$. Let $p(\alpha_j) = \epsilon$ for all $j \neq i_1, i_2$, $p(\alpha_{i_1}) = q$, and $p(\alpha_{i_2}) = 1- (K-2)\epsilon - q$. By letting $\epsilon$ be arbitrarily small, we can ensure $\prod_{j=1}^K p(x_j) \neq \prod_{j=1}^K p(y_j)$ as one of the products will have more $\epsilon$ factors. We then adjust $q$ such that $\sum_{j=1}^K \alpha_j^t p(\alpha_j) = \beta$.
            \item For all $1 \leq j \leq K$, $|\beta_{j}(x,y)|=C > 0$ for some positive integer $C$. Because $|\A| > 2$, we can choose $\alpha_{i_1}$ and $\alpha_{i_2}$ such that $|\alpha^t_{i_1} - \beta| \neq |\alpha^t_{i_2} - \beta|$. Then let $p(\alpha_j) = \epsilon$ for $j \neq i_1, i_2$, $p(\alpha_{i_1}) = q$, and $p(\alpha_{i_2}) = 1- (K-2)\epsilon - q$. The argument above still applies because
            in order for $\sum_{j=1}^K \alpha_j^t p(\alpha_j) = \beta$ to hold, we need $q \approx (\beta - \alpha_{i_2}^t)/(\alpha_{i_1}^t - \alpha_{i_2}^t)$. Since $|\alpha^t_{i_1} - \beta| \neq |\alpha^t_{i_2} - \beta|$, $1-(K-2)\epsilon -q \neq q$. Although $\prod_{j=1}^K p(x_j)$ and $\prod_{j=1}^K p(y_j)$ may have the same number of $\epsilon$ factors, $1-(K-2)\epsilon -q \neq q$ prevents the two products from being equal.
        \end{enumerate}
    \item
    There exists some $2 \leq i \leq K-1$ such that $\beta = \alpha_i^t$. If $\beta_i(x,y) \neq 0$, we can just let $p(\alpha_i) = 1-\sum_{j=1, j \neq i}^K\epsilon_j$ and $p(\alpha_j) = \epsilon_j$ for $j \neq i$. By letting $\epsilon_j$ be arbitrarily small, we can still ensure that $\sum_{k=1}^K \alpha_k^tp(\alpha_k) = \beta$ and $\prod_{k=1}^K p(x_k) \neq \prod_{k=1}^K p(y_k)$. For $x,y \in \A^n$ such that $\beta_i(x,y) = 0$, we just repeat the argument for case 1 without choosing $i_1$ or $i_2$ to be $i$, and here we need $|\A| > 3$.
    \end{enumerate}

    When $t$ is even, we can still relabel $\alpha_i$ such that $\alpha_1^t \leq \alpha_2^t \leq \cdots \leq \alpha_K^t$. With the requirement that $|\A| > 4$, we can easily generalize the argument for odd moments except for the problematic case where $\beta = \alpha_i^t$, $\beta_i(x,y) = 0$, $|\beta_j(x,y)| = C > 0$ for all $1 \leq j \leq K$ and $j \neq i$, and $|\alpha_j^t - \beta| = D > 0$ for all $j \neq i$. In this case, $|\{\alpha_i^t : 1 \leq i \leq K\}| = 3$. Since $|\A| > 4$ and $\sum_{j=1}^K \beta_j(x,y) = 0$, we can only have $\alpha_1^t = \alpha_2^t$ and $\alpha_{K-1}^t = \alpha_{K}^t$. Then we can let $p(\alpha_1) = p(\alpha_2) = (1 - \epsilon)/4$, $p(\alpha_K) = (1-\epsilon)/2 - \delta$, $p(\alpha_{K-1}) = \delta$ and $p(\alpha_i) = \epsilon$. By letting $\epsilon$ and $\delta$ be arbitrarily small, we can repeat the same argument as above.

    When $|\A|=4$, the set $\A = \{-1,0,1,\surd 2\}$ with the null of the second moment equaling one is equivalent to the null that $0$ and $\surd 2$ have equal mass. A randomization test can be constructed for this latter null by permuting these two numbers; see Example \ref{ex:construct_discrete_2} of Appendix \ref{app:examples}.
\end{proof}

\begin{proof}[of Proposition~\ref{prop:discrete_quantile}]
    Let $x,y \in \A^n$ be arbitrary such that $c(\alpha;x) > c(\alpha;y)$ for some $\alpha \in \A$ and let $\beta_i(x,y) = c(\alpha_i;x) - c(\alpha_i;y)$. We know that there exists some $1 \leq i \leq K-1$ such that $\alpha_i \leq q < \alpha_{i+1}$. One of the following two cases must occur.
    \begin{enumerate}
    \item There exists some $i_1 \leq i$ and $i_2 > i$ such that $|\beta_{i_1}(x,y)| \neq |\beta_{i_2}(x,y)|$. Then we let $p(\alpha_j) = \epsilon$ for $j \neq {i_1}, {i_2}$, $p(\alpha_{i_1}) = p - (i-1)\epsilon$, and $p(\alpha_{i_2}) = 1- (K-i-1)\epsilon - p$. One of $\prod_{j=1}^K p(x_j)$ and $\prod_{j=1}^K p(y_j)$ will have more $\epsilon$ factors and hence they are not the same.
    \item For all $1 \leq j \leq K$, $|\beta_{j}(x,y)|=C > 0$ for some constant $C$. Because $|\A| > 2$, without loss of generality, we can assume that there exists $i_1,i_2,i_3$ such that $i_1 < i_2 \leq i < i_3$. Then we let $p(\alpha_j) = \epsilon$ for $j \neq {i_1}, {i_2},{i_3}$, $p(\alpha_{i_1}) = p/2$, $p(\alpha_{i_2}) = p/2 - (i-2)\epsilon$, and $p(\alpha_{i_3}) = 1- (K-i-1)\epsilon - p$. The argument above still applies because one of $\prod_{j=1}^K p(x_j)$ and $\prod_{j=1}^K p(y_j)$ will have more $\epsilon$ factors.
    \end{enumerate}
\end{proof}

\begin{proof}[of Proposition~\ref{prop:linear}]
Since $\G$ is a group of linear maps, each element is an invertible $n\times n$ matrix $A = \{a_{ij}\}_{ij}$. Suppose there exist distinct $i,i'$ and some $j'$ such that $a_{ij'},a_{i'j'} \neq 0$. If $\Omega_A$ is non-empty, then the two coordinates $\sum_j a_{ij}X_j$ and $\sum_j a_{i'j}X_j$ of $AX$ are independent because $AX \eqd X$. We can thus apply the Darmois--Skitovich theorem \citep[for a reference, see Chapter 3 of][]{kagan1973characterization}, which implies that since $a_{ij'}a_{i'j'} \neq 0$, $X_{j'}$ is Gaussian. Since $X_1,\ldots,X_n$ are i.i.d., they are therefore all Gaussian, so $\Omega_A$ is a subset of Gaussian distributions.

Alternatively, suppose there do not exist distinct $i,i'$ and some $j'$ such that $a_{ij'},a_{i'j'} \neq 0$. Then each column of $A$ has exactly one non-zero entry. Since $A$ is invertible, each row must also have exactly one non-zero entry. Thus $A$ is of the form
\[
\begin{bmatrix}
d_{1} & 0 & \ldots & 0 \\
0 & d_2 & 0 &\vdots \\
\vdots & 0 & \ddots & 0 \\
0 & \ldots & 0 & d_n
\end{bmatrix} \sigma_n,
\]
where each $d_i \neq 0$ and $\sigma_n$ is a permutation of the coordinates.

For $\Omega_A$ to be non-empty, we must therefore have
\begin{equation}\label{eq:proof-diag}
    X_i \eqd d_j X_i
\end{equation}
for each corresponding $d_j$. This requires $|d_j|=1$. To see this, suppose first that $|d_j|>1$. Then for any $c>0$ and any positive integer $k$, \eqref{eq:proof-diag} implies
\[
    \PP(|X_i|\leq c)
    =
    \PP(|X_i|\leq c/|d_j|^k).
\]
Taking $k\rightarrow\infty$ implies
\[
    \PP(|X_i|\leq c)=\PP(X_i=0),
\]
which for all $c>0$ implies that $X_i=0$ almost surely, contradicting continuity. If $|d_j|<1$, the same argument applies to $d_j^{-1}$. Hence $d_j\in\{-1,1\}$.

If some $d_j=-1$, invariance requires $X_i \eqd -X_i$, which holds if and only if $P$ is symmetric around zero. Thus $\Omega_A$ equals the distributions symmetric around zero. If instead $d_j=1$ for every $j$, then $A$ is a permutation matrix and $\Omega_A=\Omega$. Since $\G$ is a collection of invertible linear maps, $\Omega_{\G}$ is an intersection of these possible sets of distributions, as required.
\end{proof}

\begin{proof}[of Proposition~\ref{prop:cts_case1}]
    Let $X$ be a continuous random variable with probability density function $p$ supported on $m$ intervals $[x_1, y_1], \ldots, [x_{m}, y_{m}]$. We will show that for any $\beta \in \R$ we can find an $\alpha \in (0,1]$ and a probability density function $g$ on the real line with $(2k+1)$th moment equal to $\beta$ and $g(x)|_{\bigcup_{i=1}^{m}[x_i,y_i]} = \alpha p(x)$. Let $h$ be a probability density function supported on $\R \setminus \bigcup_{i=1}^{m} [x_i,y_i]$ and let $Y$ be the random variable with probability density $h$. For any $a \in (0,1]$, let $g_a(x) = ap(x) + (1-a)h(x)$. Let $X_a$ be the random variable with probability density $g_a$. Then
    \[
    \E(X_a^{2k+1}) = a\E(X^{2k+1}) + (1-a)\E(Y^{2k+1}).
    \]
    Let $P$ be a positive number such that $P^{2k+1} > \beta$ and $Q$ be a negative number such that $Q^{2k+1} < \beta$. Let
    \[
        h(x)=
    \begin{cases}
      \ind_{[P,P+1]}, & \E(X^{2k+1}) \leq \beta, \\
      \ind_{[Q-1, Q]}, & \E(X^{2k+1}) > \beta.
    \end{cases}
    \]
    Then
    \[
        \E(Y^{2k+1})=
    \begin{cases}
      \dfrac{1}{2k+2} \sum_{i=0}^{2k+1} P^i(P+1)^{{2k+1}-i} \geq P^{2k+1} > \beta, & \E(X^{2k+1}) \leq \beta, \\[6pt]
      \dfrac{1}{2k+2} \sum_{i=0}^{2k+1} (Q-1)^iQ^{{2k+1}-i} \leq Q^{2k+1} < \beta, & \E(X^{2k+1}) > \beta.
    \end{cases}
    \]
    In both cases, we can find an $\alpha \in (0,1]$ such that $\E(X_\alpha^{2k+1}) = \beta$.
    Thus, for any $\beta \in \R$ and $k \in \mathbb{Z}_{\geq 0}$, $\{P \in \Omc(\R) : \E_P(X_i^{2k+1}) = \beta\}$ is locally dense.

    We now consider the $2k$th moment, and thus require that $\beta > 0$, for otherwise the null hypothesis is empty. Let $X$ be a continuous random variable with probability density function $p$ supported on $m$ intervals $[x_1, y_1], \ldots, [x_{m}, y_{m}]$ where $\max_{1 \leq i \leq m}|y_i-x_i| < \beta^{1/(2k)}/(m+1)$. We can find an $\alpha \in (0,1]$ and a probability density function $g$ on the real line with $2k$th moment equal to $\beta$ and $g(x)|_{\bigcup_{i=1}^{m}[x_i,y_i]} = \alpha p(x)$. The choice of the upper bound for the interval length plays a crucial role in the definition of local denseness. If we remove that condition from the definition, then we can choose a probability density function $p(x)$ to be supported on $\R \setminus [-2\beta^{1/(2k)}, 2\beta^{1/(2k)}]$. In this case, whatever probability density function $g$ and $\alpha \in (0,1]$ we choose such that $g(x)|_{\bigcup_{i=1}^{m}[x_i,y_i]} = \alpha p(x)$, the $2k$th moment of the random variable with density $g$ is strictly greater than $\beta$.

    Again, let $h$ be a probability density function supported on $\R \setminus \bigcup_{i=1}^{m} [x_i,y_i]$ and let $Y$ be the random variable with probability density $h$. For any $a \in (0,1]$, let $g_a(x) = ap(x) + (1-a)h(x)$. Let $X_a$ be the random variable with probability density $g_a$. Then
    \[
    \E(X_a^{2k}) = a\E(X^{2k}) + (1-a)\E(Y^{2k}).
    \]
    If $\E(X^{2k}) \leq \beta$, let $P$ be a positive number such that $P^{2k} > \beta$ and $[P, P + 1] \subset \R \setminus \bigcup_{i=1}^{m} [x_i,y_i]$ and let
    \[
    h(x) = \ind_{[P,P+1]}.
    \]
    Then $\E(Y^{2k}) \geq P^{2k} > \beta$, and we can easily find an $\alpha \in (0,1]$ such that $\E(X^{2k}_\alpha) = \beta$.

    On the other hand, if $\E(X^{2k}) > \beta$, given the condition $\max_{1 \leq i \leq m}|y_i-x_i| < \beta^{1/(2k)}/(m+1)$, we can find an interval $[x,y] \subset [0,\beta^{1/(2k)}]$ such that $[x,y] \cap \bigcup_{i=1}^m[x_i,y_i] = \emptyset$.
    Let
    \[
    h(x) = \frac{1}{y-x}\ind_{[x,y]}.
    \]
    Then $\E(Y^{2k}) \leq y^{2k} \leq \beta$.
    We can then find an $\alpha \in (0,1]$ such that $\E(X^{2k}_\alpha) = \beta$.

    Thus, for any $\beta \in \R_{>0}$ and $k \in \mathbb{Z}_{\geq 1}$, $\{P \in \Omc(\R) : \E_P(X_i^{2k}) = \beta\}$ is locally dense.
\end{proof}

\begin{proof}[of Proposition~\ref{prop:cts_case2}]
Let $X$ be a continuous random variable with probability density function $f$ supported on $m$ intervals $[x_1, y_1], \ldots, [x_{m}, y_{m}]$. We will show that for any $p \in (0,1)$ and $q \in \R$, we can find an $\alpha \in (0,1]$ and a probability density function $g$ on the real line with the $p$th quantile equal to $q$ and $g(x)|_{\bigcup_{i=1}^{m}[x_i,y_i]} = \alpha f(x)$.

Let $h$ be a probability density function supported on $\R \setminus \bigcup_{i=1}^{m} [x_i,y_i]$ and let $Y$ be the random variable with probability density $h$. For any $a \in (0,1]$, let $g_a(x) = af(x) + (1-a)h(x)$. Let $X_a$ be the random variable with probability density $g_a$. Then
\[
\PP(X_a \leq q) = a\,\PP(X \leq q) + (1-a)\,\PP(Y \leq q).
\]
There exists a large integer $N$ such that $[x_1, y_1], \ldots, [x_m,y_m], q \subset [-N+1,N-1]$. Choose $\alpha >0$ small enough such that $\alpha\,\PP(X \leq q) < p$. In order for $\PP(X_\alpha \leq q) = p$, we need $\PP(Y \leq q) = (p- \alpha\,\PP(X\leq q))/(1-\alpha)$. This can be achieved by letting
\[
h(x) = \frac{p- \alpha\,\PP(X\leq q)}{1-\alpha}\ind_{[-N-1,-N]} + \left(1-\frac{p- \alpha\,\PP(X\leq q)}{1-\alpha}\right)\ind_{[N,N+1]}.
\]
\end{proof}

\subsection{Proof of Proposition~\ref{prop:continuous}}\label{app:proofs2}

We prove Proposition~\ref{prop:continuous} by making use of Proposition~\ref{prop:tool} and showing that if $\Omega_0$ is locally dense, then there is no function that is invariant to all distributions in $\Omega_0$ that is also not invariant to all distributions in $\Omega$.
To do so, we partition the space of functions into three classes of functions.
The first are approximate permutation maps, which behave like permutation maps.
We will show that these functions are invariant to all continuous distributions, so cannot be used to construct a randomization test.
The other two classes of functions will satisfy properties that allow us to construct, for any such function, a distribution in $\Omega_0$ that is not invariant to the function, which would yield the result.

We begin by defining some notation and concepts. For a subset $A \subseteq \R^n$, we let $m(A)$ denote the Lebesgue measure of the set. We let $\sigma_1,\ldots,\sigma_{n!}$ be the distinct $n!$ permutations of coordinates. Let $A_i = \{\sigma_i(x): x \in A\}$ be the image of $A$ under the permutation map $\sigma_i$. Let $\bar{A} = \cup_{i=1}^{n!} A_i$ be the union of all the images of $A$ under permutation maps.

\begin{definition}
    We call a bijective function $f: \R^n \rightarrow \R^n$ mass-preserving if for any measurable set $A \subset \R^n$, $m(A)=0$ if and only if $m(f(A)) = 0$.
\end{definition}

\begin{definition}
    A bijective function $f: \R^n \rightarrow \R^n$ is orbit-preserving if for any measurable set $A \subset \R^n$, $m(f(A) \setminus \bar{A}) = 0$.
\end{definition}

\begin{definition}
    A bijective function $f: \R^n \rightarrow \R^n$ is box-orbit-preserving if for any box $A = \prod_{i=1}^n [a_i,b_i] \subset \R^n$, $m(f(A) \setminus \bar{A}) = 0$.
\end{definition}

\begin{definition}
    We call a bijective function $f: \R^n \rightarrow \R^n$ an approximate permutation map if $f$ is box-orbit-preserving and mass-preserving.
\end{definition}

Both the orbit-preserving and box-orbit-preserving properties imply that for any measurable set $A \subset \R^n$, the images of $A$ under the approximate permutation maps are `contained' in the images of $A$ under the actual permutation maps, not the other way around. However, it suffices to require only the box-orbit-preserving and mass-preserving properties to define approximate permutation maps because these properties apply to all the measurable sets $A$, and this turns out to be exactly as strong as we need.

A function is either an approximate permutation map, or not mass-preserving, or mass-preserving but not box-orbit-preserving.
The proof of Proposition~\ref{prop:continuous} will follow by showing that functions that are approximate permutation maps are invariant to all distributions, while functions in the latter two classes must always fail to be invariant to some distribution in $\Omega_0$ when it is locally dense.
We consider each space of functions separately in the following series of lemmas, which will immediately obtain the main result.
The proofs of the lemmas are provided at the end of this appendix so as not to disrupt the flow of the arguments here.

\begin{lemma}\label{lemma1}
    A box-orbit-preserving map is orbit-preserving.
\end{lemma}

\begin{lemma}\label{lemma2}
    An approximate permutation map is invariant to all i.i.d. continuous distributions.
\end{lemma}

We next consider functions that are not approximate permutation maps. These functions are either not mass-preserving or mass-preserving but not box-orbit-preserving.
For such functions, we will essentially find several small intervals on which the map $f$ acts very differently from the permutation maps. On those intervals, we can define distributions in $\Omega_0$ that are not preserved by $f$. Outside those intervals, we will define distributions based on the local denseness property. Consequently, when we combine these distributions, we will have an $f$-specific distribution that is not invariant to $f$ and belongs to $\Omega_0$.

\begin{lemma}\label{lemma3}
For any function $f: \R^n \rightarrow \R^n$ that is not mass-preserving, there exists a random vector $X = (X_1,\ldots,X_n)$, with $X_i$ drawn i.i.d. from distribution $P \in \Omega_0$ where $\Omega_0$ is locally dense, such that $X$ is not invariant under $f$.
\end{lemma}

\begin{lemma}\label{lemma4}
    For any mass-preserving and not box-orbit-preserving function $f: \R^n \rightarrow \R^n$, there exists a random vector $X = (X_1,\ldots,X_n)$, with $X_i$ drawn i.i.d. from distribution $P \in \Omega_0$ where $\Omega_0$ is locally dense, such that $X$ is not invariant under $f$.
\end{lemma}

The proof of Proposition~\ref{prop:continuous} follows immediately from the above lemmas.

\begin{proof}[of Proposition~\ref{prop:continuous}]
By definition, a function is either an approximate permutation map, not mass-preserving, or mass-preserving but not box-orbit-preserving. If it is an approximate permutation map, Lemma~\ref{lemma1} shows it is invariant to all distributions. If it is not, Lemmas~\ref{lemma2}--\ref{lemma4} show it is not invariant to all distributions in $\Omega_0$ when it is locally dense. The result then follows by Proposition~\ref{prop:tool}.
\end{proof}

We now prove the lemmas. 

\begin{proof}[of Lemma~\ref{lemma1}]
Assume for contradiction that $f$ is box-orbit-preserving and not orbit-preserving. Then there exists a measurable set $A \subset \R^n$ such that $m(f(A) \setminus \bar{A}) = c > 0$. Let $B = f(A) \setminus \bar{A}$, so that $B \cap \overline{f^{-1}(B)} = \emptyset$. Since $f$ is mass-preserving, $m(f^{-1}(B)) > 0$. We can find a finite number of pairwise disjoint boxes $D_1, \ldots, D_k$ where $D_j = \prod_{i=1}^{n} [a^j_i,b^j_i]$ such that $f^{-1}(B) \subset \cup_{j=1}^k D_j$ and $m(\cup_{j=1}^k D_j \setminus f^{-1}(B)) < c/(2n!)$. Let $D = \cup_{j=1}^k D_j$. Then
\[
    m(f(D) \setminus \bar{D}) \geq m(B \setminus \bar{D}) = m(B\setminus \overline{(D \setminus f^{-1}(B))}) \geq m(B) - n!\,m(D \setminus f^{-1}(B)) \geq c/2 > 0.
\]
Thus, there exists at least one $D_j$ such that $m(f(D_j) \setminus \bar{D_j}) > 0$.
\end{proof}

\begin{proof}[of Lemma~\ref{lemma2}]
    First, for any measurable set $A \subset \R^n$, we can find a disjoint partition of $f(A) = \bigsqcup_{i=1}^{n!} B_i \bigsqcup C$ such that $B_i \subset \sigma_i(A)$, $C \subset \bar{A}^c$. By Lemma~\ref{lemma1}, we know that an approximate permutation map is also orbit-preserving. Thus, $m(f(A) \setminus \bar{A}) = 0$ and $m(C) = 0$. Let $A_i = f^{-1}(B_i)$. We would like to show that $\sigma_i (A_i) = B_i$ almost everywhere, that is, $m(\sigma_i (A_i) \triangle B_i) = 0$, where $\sigma_i (A_i) \triangle B_i$ denotes the symmetric difference between the two sets $\sigma_i(A_i)$ and $B_i$. Then, it follows that
    \[
        \PP(X \in f(A)) = \sum_{i=1}^{n!}\PP(X \in B_i)=\sum_{i=1}^{n!} \PP(X \in \sigma_i(A_i))= \sum_{i=1}^{n!} \PP(X \in A_i)= \PP(X \in A).
    \]

    Now to show that $m(\sigma_i (A_i) \triangle B_i) = 0$, we need to show $m(\sigma_i (A_i) \setminus B_i) = 0$ and $m(B_i \setminus \sigma_i (A_i))= 0$. The latter is trivial as $m(B_i \setminus \sigma_i (A_i)) = m(f(A_i) \setminus \sigma_i(A_i)) = 0$ follows from the condition given. It is only left to show that $m(\sigma_i (A_i) \setminus B_i) = m(\sigma_i(A_i) \setminus f(A_i)) = 0$.

    Assume for contradiction that there exists an $i$ where $m(\sigma_i(A_i) \setminus f(A_i)) > 0$. Together with $m(f(A_i) \setminus \sigma_i(A_i)) = 0$, this implies that $m(A_i) = m(\sigma_i(A_i)) > m(f(A_i))$. Since we can decompose $\bar{A}$ in terms of disjoint unions of $A_j$ where $m(A_j) \geq m(f(A_j))$, it implies that $m(\bar{A}) > m(f(\bar{A}))$ and $m(\bar{A} \setminus f(\bar{A})) > 0$.

    However, let $C = \bar{A} \setminus f(\bar{A})$. Since $\bar{C} \subset \bar{A}$ and $f^{-1}(C) \cap \bar{A} = \emptyset$, we have $f^{-1}(C) \cap \bar{C} = \emptyset$. Let $D = f^{-1}(C)$. Because $D \cap \bar{C} = \emptyset$, we have $\bar{D} \cap \bar{C} = \emptyset$. Then, $m(f(D) \setminus \bar{D}) = m(C \setminus \bar{D}) = m(C) > 0$. This is a contradiction.
\end{proof}

\begin{proof}[of Lemma~\ref{lemma3}]
    Because $f$ is not mass-preserving, there exists a measurable set $A \subset \R^n$ such that $m(A) = 0$ and $m(f(A))>0$. Thus, we can always find an arbitrarily small closed box $C = \prod_{i=1}^n [x_i,y_i]$ where $[x_i,y_i]$ and $[x_j,y_j]$ are either disjoint or equal such that $m(C \cap f(A)) > 0$.

    Consider the probability density function
    \[
    p(x) = \sum_{i=1}^n \frac{\ind_{[x_i,y_i]}}{n|y_i-x_i|}.
    \]
    By the definition of local denseness, we can find an $\alpha \in (0,1]$ and a distribution $G \in \Omega_0$ with density function $g$ such that $g(x)|_{\bigcup_{i=1}^{n}[x_i,y_i]} = \alpha p(x)$. Let $X_1,\ldots,X_n$ be random variables with probability density $g$.

    Let $X = (X_1,\ldots,X_n)$. Then
    \[
    \PP(X \in C \cap f(A)) = \frac{m(C \cap f(A))}{m(C)}\,\PP(X \in C) > 0 = \PP(X \in f^{-1}(C) \cap A),
    \]
    where the last equality is due to the fact that $m(A) = 0$. Clearly, $X$ is not invariant under $f$.
\end{proof}

\begin{proof}[of Lemma~\ref{lemma4}]
Let $n$ be a positive integer. A partition of $n$ is a sequence $\lambda = (\lambda_1, \lambda_2,\ldots)$ of non-negative integers in decreasing order $\lambda_1 \geq \lambda_2 \geq \cdots$
such that $\lambda_1 + \lambda_2 + \cdots = n$. Let $P_n$ denote the set of all partitions of $n$. We will use $L_n$ to denote the reverse lexicographic ordering on the set $P_n$, where $L_n$ is the subset of $P_n \times P_n$ consisting of all $(\lambda, \mu)$ such that either $\lambda = \mu$ or the first non-vanishing difference $\lambda_i - \mu_i$ is positive. The relation $L_n$ is a total ordering.
We prove the lemma by induction on the reverse lexicographic ordering on partitions of $n$.

Since $f$ is not box-orbit-preserving, there exists a box $A = \prod_{i=1}^n [u_i, v_i]$ such that $m(f(A) \setminus \bar{A}) > 0$.
Partition $A$ into finitely many disjoint smaller boxes of the form $\prod_{i=1}^n [x_i, y_i]$, where each interval $[x_i, y_i]$ is either disjoint from or equal to another. Since $m(f(A) \setminus \bar{A}) > 0$, there exists at least one such box $B$ with $m(f(B) \setminus \bar{B}) > 0$.

Each such box $B = \prod_{i=1}^n [x_i, y_i]$ induces a partition $\lambda \in P_n$ by recording the multiplicities of the intervals $[x_i, y_i]$. We proceed by induction on $P_n$ under the reverse lexicographic order.

\textit{Base case}. Suppose $\lambda = (n)$, so $B = [x_1, y_1]^n$. Let $C = f(B) \setminus \bar{B}$. Since $m(C) > 0$ and $f$ is mass-preserving, $m(f^{-1}(C)) > 0$.
Consider the density
\[
p(x) = \frac{1}{|y_1 - x_1|} \ind_{[x_1, y_1]}(x).
\]
By the local denseness property, there exist $\alpha \in (0,1]$ and a distribution $G \in \Omega_0$ with density $g$ such that $g(x) = \alpha p(x)$ on $[x_1, y_1]$. Let $X = (X_1,\ldots,X_n)$ where the $X_i$ are drawn i.i.d. from $G$. Then
\[
\PP(X \in f^{-1}(C)) = \frac{m(f^{-1}(C))}{m(B)} \alpha^n > 0,
\quad
\PP(X \in C) = 0,
\]
so $X$ is not invariant under $f$.

\textit{Inductive step}. Assume the statement holds for all partitions $\mu \succ \lambda$.
Let $B = \prod_{i=1}^n [x_i, y_i]$ correspond to $\lambda$, and rename the intervals so that $[a_i, b_i]$ appears $\lambda_i$ times. Define the set
\[
S = \left\{ D = \prod_{i=1}^n [w_i, z_i] : [w_i, z_i] \subset \{ [a_1, b_1],\ldots, [a_m, b_m] \} \right\} \setminus \bar{B},
\]
and let $S_{\leq} \subset S$ denote those boxes corresponding to partitions less than or equal to $\lambda$ and let $S_{>} \subset S$ denote those boxes corresponding to partitions larger than $\lambda$.
We consider three cases.

\textit{Case} 1. Suppose there exists $D \in S_>$ such that $m(f(B) \cap D) > 0$.
Then by the induction hypothesis, there exists a distribution $X$, with i.i.d. coordinates from a distribution in $\Omega_0$, such that $X$ is not invariant under $f$. This is because $\bar{B} \cap \bar{D} = \emptyset$ and $m(f^{-1}(D) \setminus \bar{D}) \geq m(f^{-1}(D) \cap B) > 0$.

\textit{Case} 2. Suppose there exists $D \in S_{\leq}$ such that $m(f(B) \cap D) > 0$. Let $\mu_i$ denote the number of times $[a_i,b_i]$ appears in $D$. The vector $(\mu_1,\ldots,\mu_m)$ might not be a partition, but the first non-vanishing difference of $\lambda_i - \mu_i$ must be positive. Let $j$ be the smallest integer such that $\lambda_j > \mu_j$. Define the density
\[
p(x) = \sum_{i=1}^m (\epsilon_{i-1} - \epsilon_i) \frac{\ind_{[a_i, b_i]}(x)}{|b_i - a_i|},
\]
where $\epsilon_0 = 1$, $\epsilon_m = 0$, and the $\epsilon_i$ are chosen later.
By the local denseness property, there exist $\alpha \in (0,1]$ and $G \in \Omega_0$ with density $g$ such that $g(x) = \alpha p(x)$ on $\bigcup_i [a_i, b_i]$. Let $X = (X_1,\ldots,X_n)$ be i.i.d. samples from $G$.
Let $E = f(B) \cap D$. Then $m(f^{-1}(E)) > 0$, and
\[
\PP(X \in E) \leq \PP(X \in D) = \alpha^n \prod_{i=1}^m (\epsilon_{i-1} - \epsilon_i)^{\mu_i},
\qquad
\PP(X \in f^{-1}(E)) = \frac{m(f^{-1}(E))}{m(B)} \alpha^n \prod_{i=1}^m (\epsilon_{i-1} - \epsilon_i)^{\lambda_i}.
\]
We choose $\epsilon_{i-1} - \epsilon_i = \delta$ small for all $i > j$, and choose $\epsilon_{j-1} - \epsilon_j$ so that
\[
\epsilon_{j-1} - \epsilon_j > \frac{m(B)}{m(f^{-1}(E))} \delta.
\]
Thus,
\[
\frac{\PP(X \in f^{-1}(E))}{\PP(X \in E)} \geq \left( \frac{\epsilon_{j-1} - \epsilon_j}{\delta} \right)^{\lambda_j - \mu_j} \frac{m(f^{-1}(E))}{m(B)} > 1,
\]
so $X$ is not invariant under $f$.

\textit{Case} 3. Suppose $m(f(B) \cap S^c \cap \bar{B}^c) > 0$.
Define
\[
p(x) = \sum_{i=1}^m \frac{1}{m |a_i - b_i|} \ind_{[a_i, b_i]}(x),
\]
and apply the local denseness property to find $\alpha \in (0,1]$ and $G \in \Omega_0$ such that $g(x) = \alpha p(x)$ on $\bigcup_i [a_i, b_i]$. Let $X = (X_1,\ldots,X_n)$ be i.i.d. samples from $G$. Let $F = f(B) \cap S^c \cap \bar{B}^c$. Then
\[
\PP(X \in F) \leq \PP(X \in S^c \cap \bar{B}^c) = 0,
\]
but
\[
\PP(X \in f^{-1}(F)) = \frac{m(f^{-1}(F))}{m(B)}\, \PP(X \in B) > 0,
\]
so again $X$ is not invariant under $f$. This concludes the proof.
\end{proof}

\subsection{Statements and proofs of claims made in Remark \ref{rm:extend-cts}}\label{app:proofs3}

This subsection states and proves the claims made in Remark \ref{rm:extend-cts}.

\begin{proposition}\label{prop:cts_subset}
    The statement in Proposition~\ref{prop:cts_case1} holds if we replace $\Omc(\R)$ with $\Omc([b_0,b_1])$ for some interval $[b_0,b_1]$ and $\beta$ in the support. The same holds for Proposition~\ref{prop:cts_case2} for any $q \in (b_0,b_1)$.
\end{proposition}

\begin{proof}[of Proposition~\ref{prop:cts_subset}]
We first consider the statement in Proposition~\ref{prop:cts_case1} replacing $\Omc(\R)$ with $\Omc([b_0,b_1])$ for some interval $[b_0,b_1]$.
We will separately consider the cases where the moment order is odd and even.

For the case where the moment order is odd, consider any positive integer $k$, and consider $\beta$ such that $b_0^{2k+1}<\beta<b_1^{2k+1}$. For any $m$, we will show that the null is locally dense by considering intervals $[x_1, y_1], \ldots, [x_{m}, y_{m}] \subset \mathcal{B}$ where
\[
\max_{1 \leq i \leq m}|y_i-x_i| < \frac{\min\{b_1-\beta^{1/(2k+1)},\ \beta^{1/(2k+1)} - b_0\}}{m+1}.
\]
Consider any $X$ with probability density $p$ on such intervals. We want to show that we can find an $\alpha \in (0,1]$ and a probability density function $g$ on $\mathcal{B}$ with $(2k+1)$th moment equal to $\beta$ and $g(x)|_{\bigcup_{i=1}^{m}[x_i,y_i]} = \alpha p(x)$.
To do so, let $h$ be a probability density function supported on $\mathcal{B} \setminus \bigcup_{i=1}^{m} [x_i,y_i]$ and let $Y$ be the random variable with probability density $h$. For any $a \in (0,1]$, let $g_a(x) = ap(x) + (1-a)h(x)$. Let $X_a$ be the random variable with probability density $g_a$. Then
\[
\E(X_a^{2k + 1}) = a\E(X^{2k+1}) + (1-a)\E(Y^{2k+1}).
\]
If $\E(X^{2k+1}) \geq \beta$, given the condition $\max_{1 \leq i \leq m}|y_i-x_i| < (\beta^{1/(2k+1)}-b_0)/(m+1)$, we can find an interval $[x,y] \subset [b_0,\beta^{1/(2k+1)}]$ such that $[x,y] \cap \bigcup_{i=1}^m[x_i,y_i] = \emptyset$.
Let
\[
h(x) = \frac{1}{y-x}\ind_{[x,y]}.
\]
Then $\E(Y^{2k + 1}) \leq y^{2k+1} \leq \beta$.
We can then find an $\alpha \in (0,1]$ such that $\E(X^{2k+1}_\alpha) = \beta$.

Similarly, if $\E(X^{2k+1}) \leq \beta$, given the condition $\max_{1 \leq i \leq m}|y_i-x_i| < (b_1-\beta^{1/(2k+1)})/(m+1)$, we can find an interval $[x,y] \subset [\beta^{1/(2k+1)},b_1]$ such that $[x,y] \cap \bigcup_{i=1}^m[x_i,y_i] = \emptyset$.
Let
\[
h(x) = \frac{1}{y-x}\ind_{[x,y]}.
\]
Then $\E(Y^{2k + 1}) \geq x^{2k+1} \geq \beta$.
We can then find an $\alpha \in (0,1]$ such that $\E(X^{2k+1}_\alpha) = \beta$.

For the case where the moment order is even, we will show that for any real number $0<\beta< \max \{b_0^{2k},b_1^{2k}\}$ and any continuous random variable $X$ with probability density function $p$ supported on $m$ intervals $[x_1, y_1], \ldots, [x_{m}, y_{m}] \subset \mathcal{B}$ where
\[
\max_{1 \leq i \leq m}|y_i-x_i| < \frac{\min\{\max\{|b_0|,|b_1|\}-\beta^{1/(2k)},\ \beta^{1/(2k)}\}}{m+1},
\]
we can find an $\alpha \in (0,1]$ and a probability density function $g$ on $\mathcal{B}$ with $2k$th moment equal to $\beta$ and $g(x)|_{\bigcup_{i=1}^{m}[x_i,y_i]} = \alpha p(x)$.

To show this, let us assume without loss of generality that $|b_1| \geq |b_0|$. Let $h$ be a probability density function supported on $\mathcal{B} \setminus \bigcup_{i=1}^{m} [x_i,y_i]$ and let $Y$ be the random variable with probability density $h$. For any $a \in (0,1]$, let $g_a(x) = ap(x) + (1-a)h(x)$. Let $X_a$ be the random variable with probability density $g_a$. Then
\[
\E(X_a^{2k}) = a\E(X^{2k}) + (1-a)\E(Y^{2k}).
\]
If $\E(X^{2k}) \geq \beta$, given the condition $\max_{1 \leq i \leq m}|y_i-x_i| < \beta^{1/(2k)}/(m+1)$, we can find an interval $[x,y] \subset [-\beta^{1/(2k)},\beta^{1/(2k)}]$ such that $[x,y] \cap \bigcup_{i=1}^m[x_i,y_i] = \emptyset$.
Let
\[
h(x) = \frac{1}{y-x}\ind_{[x,y]}.
\]
Then $\E(Y^{2k}) \leq y^{2k} < \beta$.
We can then find an $\alpha \in (0,1]$ such that $\E(X^{2k}_\alpha) = \beta$.

Similarly, if $\E(X^{2k}) \leq \beta$, given the condition $\max_{1 \leq i \leq m}|y_i-x_i| < (|b_1|-\beta^{1/(2k)})/(m+1)$, we can find an interval $[x,y] \subset [\beta^{1/(2k)}, \ind[b_1 > 0]\times b_1 -\ind[b_1 < 0]\times b_1]$ such that $[x,y] \cap \bigcup_{i=1}^m[x_i,y_i] = \emptyset$.
Let
\[
h(x) = \frac{1}{y-x}\ind_{[x,y]}.
\]
Then $\E(Y^{2k}) \geq x^{2k} \geq \beta$.
We can then find an $\alpha \in (0,1]$ such that $\E(X^{2k}_\alpha) = \beta$.

We next consider the statement in Proposition~\ref{prop:cts_case2} replacing $\Omc(\R)$ with $\Omc([b_0,b_1])$ for some interval $[b_0,b_1]$ and considering any $q \in (b_0,b_1)$.
Let $X$ be a continuous random variable with probability density function $f$ supported on $m$ intervals $[x_1, y_1], \ldots, [x_{m}, y_{m}] \subset [b_0,b_1]$ where $\max_{1 \leq i \leq m}|y_i-x_i| < \min\{(q-b_0)/(m+1),\ (b_1-q)/(m+1)\}$.
We will show that for any $p \in (0,1)$, there exist $\alpha \in (0,1]$ and a probability density function $g$ on $[b_0,b_1]$ such that $q$ is the $p$th quantile of $g$ and $g(x)|_{\bigcup_{i=1}^{m}[x_i,y_i]} = \alpha f(x)$.

Let $h$ be a probability density function supported on $[b_0,b_1] \setminus \bigcup_{i=1}^{m} [x_i,y_i]$ and let $Y$ be the random variable with probability density $h$. For any $a \in (0,1]$, let $g_a(x) = af(x) + (1-a)h(x)$. Let $X_a$ be the random variable with probability density $g_a$. Then
\[
\PP(X_a \leq q) = a\,\PP(X \leq q) + (1-a)\,\PP(Y \leq q).
\]
Given the condition $\max_{1 \leq i \leq m}|y_i-x_i| < \min\{(q-b_0)/(m+1),\ (b_1-q)/(m+1)\}$, we can find intervals $[x,y] \subset [b_0,q]$ and $[w,z] \subset (q,b_1]$ such that $[x,y] \cap \bigcup_{i=1}^m[x_i,y_i] = [w,z] \cap \bigcup_{i=1}^m[x_i,y_i]= \emptyset$.

Now we can choose $\alpha >0$ small enough such that $\alpha\,\PP(X \leq q) <p$. In order for $\PP(X_\alpha \leq q) = p$, we need $\PP(Y \leq q) = (p- \alpha\,\PP(X\leq q))/(1-\alpha)$. This can be achieved by letting
\[
h(x) = \frac{p- \alpha\,\PP(X\leq q)}{(1-\alpha)(y-x)}\ind_{[x,y]} + \left(1-\frac{p- \alpha\,\PP(X\leq q)}{1-\alpha}\right)\frac{1}{z-w}\ind_{[w,z]}.
\]
\end{proof}

\begin{proposition}\label{prop:cts_case3}
    For any $\beta \in \R_{>0}$, $\Omega_0 = \{P \in \Omc(\R) : \Var_P(X_i) = \beta\}$ is locally dense and thus does not satisfy the randomization hypothesis.
\end{proposition}

\begin{proof}[of Proposition~\ref{prop:cts_case3}]
    Let $X$ be a continuous random variable with probability density function $p$ supported on $m$ intervals $[x_1, y_1], \ldots, [x_{m}, y_{m}]$. For any $\beta \in \R_{> 0}$, we will show that we can find an $\alpha \in (0,1]$ and a probability density function $g$ on the real line with variance $\beta$ such that $g(x)|_{\bigcup_{i=1}^{m}[x_i,y_i]} = \alpha p(x)$.

    Let $h$ be a probability density function supported on $\R \setminus \bigcup_{i=1}^{m} [x_i,y_i]$ and let $Y$ be the random variable with probability density $h$. For any $a \in (0,1]$, let $g_a(x) = ap(x) + (1-a)h(x)$. Let $X_a$ be the random variable with probability density $g_a$. Then
    \[
    \begin{split}
        \Var(X_a)
        &= a(\Var(X) + \E(X)^2) + (1-a)(\Var(Y) + \E(Y)^2) - (a\E(X) + (1-a)\E(Y))^2\\
        &= a[(1-a)\E(X)^2 + \Var(X) + (1-a)\E(Y)^2 - 2(1-a)\E(X)\E(Y)] + (1-a)\Var(Y).
    \end{split}
    \]
    Choose $M > 0$ such that $\bigcup_{i=1}^{m} [x_i,y_i] \subset [-M,M]$. For any $\beta>0$ and $a = 1/2$, we can find $z > M+1$ such that
    \[
    a[(1-a)\E(X)^2 + \Var(X) + (1-a)z^2 - 2(1-a)z\E(X)] > \beta.
    \]
    We then choose $w \in (0,1)$ such that $w^2/12 < \beta$. Let $h(x) = w^{-1} \ind_{[z-w/2,\, z+ w/2]}$. In this case, since $\Var(Y) \geq 0$, we have $\Var(X_{1/2}) > \beta$. On the other hand, $\Var(X_0) = \Var(Y) = w^2/12 < \beta$. Because $\Var(X_a)$ is continuous in $a$, we can find an $\alpha \in (0,1/2)$ such that $\Var(X_\alpha) = \beta$. Thus, for any $\beta \in \R_{>0}$, $\{P \in \Omc(\R) : \Var_P(X_i) = \beta\}$ is locally dense.
\end{proof}

\end{document}